\documentclass[12pt,english,Preprint, aps]{revtex4-1}
\usepackage[T1]{fontenc}
\usepackage[latin9]{inputenc}
\usepackage[a4paper]{geometry}
\usepackage{textcomp}
\usepackage{amsmath}
\usepackage{amssymb}
\usepackage{esint}


\usepackage{babel}
\begin{document}

\preprint{APS/123-QED}

\title{Extension and Applications of a Variational Approach with Deformed
Derivatives}

\thanks{The authors express their gratitude to FAPERJ-Rio de
Janeiro and CNPq-Brazil for the partial financial support.}

\author{J. Weberszpil}

\email{josewebe@gmail.com}

\selectlanguage{english}%

\affiliation{Programa de Pós Graduação em Modelagem Matemática e Computação-PPGMMC,
Universidade Federal Rural do Rio de Janeiro, UFRRJ-IM/DTL\\
 Av. Governador Roberto Silveira n/n- Nova Iguaçú, Rio de Janeiro,
Brasil, 695014.}

\author{J. A. Helayël-Neto}

\email{helayel@cbpf.br}

\selectlanguage{english}%

\address{Centro Brasileiro de Pesquisas Físicas-CBPF-Rua Dr Xavier Sigaud
150,\\
 22290-180, Rio de Janeiro RJ Brasil. }

\date{\today}
\begin{abstract}
We have recently presented an extension of the standard variational
calculus to include the presence of deformed derivatives in
the Lagrangian of a system of particles and in the Lagrangian density 
of field-theoretic models. Classical Euler-Lagrange equations and the Hamiltonian 
formalism have been re-assessed in this approach. Whenever applied to a number of 
physical systems, the resulting dynamical equations come out to be the correct 
ones found in the literature, specially with mass-dependent and with non-linear
equations for classical and quantum-mechanical systems. In the present
contribution, we extend the variational approach with the intervalar
form of deformed derivatives to study higher-order dissipative systems,
with application to concrete situations, such as an accelerated
point charge - this is the problem of the Abraham-Lorentz-Dirac force - to stochastic dynamics
like the Langevin, the\textbf{ }advection-convection-reaction and
Fokker-Planck equations, Korteweg\textendash de Vries equation, Landau-Lifshits-Gilbert
equation and the Caldirola-Kanai Hamiltonian. By considering these different
applications, we show that the formulation investigated in this paper may be
a simple and promising path for dealing with dissipative,
non-linear and stochastic systems through the variational approach.
\end{abstract}


\maketitle

\section{Introduction}

In this contribution, we suggest the possibility that, with the same
variational formalism, it is possible to propose Lagrangians that
provide the equations describing the dynamics for diverse systems,
including dissipative systems, non-linear systems, and the description
of stochastic processes. By adopting the intervalar form of deformed
derivatives embedded into the Lagrangian, we show that our formalism
yields the correct equations of motion of such systems.

Here, we apply the least action principle for dissipative systems with
the use of intervalar form of deformed derivatives, without the necessity
of classical Rayleigh-Lagrange equations, that are known to account
for resistive forces linear in the velocities.

Although there are proposals for methods to study non-conservative
and dissipative systems \citep{Galley,RW}, we claim that the approach
with deformed derivatives is a simple and efficient option to obtain
the equations describing the dynamics for a broad variety of
linear, non-linear and stochastic systems.

The main purpose of this paper is to present the developments that
go beyond the issue of Lagrangian mechanics in the classical sense, but
reach the fields of dissipative, stochastic and non-linear systems.
To achieve these goals, we extend the contents of our previous
work of Ref. \citep{Nosso variacional} and extend the variational
calculus with deformed derivatives embedded into the Lagrangian to
consider higher-order derivatives. By using the intervalar form of
deformed conformable derivatives, we obtain the Euler-Lagrange (E-L)
equations for each case, showing the strong concordance with the literature.

By presenting some different cases in different areas, instead of
going deeper inside the solutions for each system, we suggest that with
the formalism presented here we can obtain the dynamical equations
that describe different physical systems. With this purpose, we apply our formulation, 
as exemplification, to the problem of the accelerated point charge - the Abraham-Lorentz-Dirac
force, to stochastic dynamics like the Langevin, the advection-convection-reaction
and Fokker-Planck(FP) equations, Korteweg-de Vries (KDV) equation, the
Landau-Lifshits-Gilbert (LLG) problem and the Caldirola-Kanai (KK) Hamiltonian.

Concerning the dynamics of stochastic systems, we show that our approach
does not need the Nelson-Yasue\textquoteright stochastic variational
method nor Itô's stochastic calculus and neither has the necessity
of heuristics auxiliary fields, in such a way that we can use our
variational approach to describe classical and quantum behaviors in
broad rage of situations \citep{Koide-Kodama-KaZuo}.

Also, it is worth to note that with our approach, the Lagrangian function
does not need to be duplicated in order to describe the coupling of the
dynamics with an additional process \citep{Massimiliano}. 

The justifications for the use of the deformed derivatives is inter-connected
to the different degrees of freedom and also related to the reversibility/irreversibility
process; some justifying details can be found in some previous
publications in Ref. \citep{EPL Weber,Jap-Weber-Sotolongo,Nosso On conection2015,Nosso variacional}.

Our paper is outlined as follows. Section 2 addresses some mathematical
aspects and justification for the use of deformed derivative . In Section
3, we focus on the extension of the variational formulations. Section
4 addresses the case of relativistic fields. In Section 5, we apply our formalism
to different systems. Finally, in Section 6, we cast our general conclusions
and possible paths for further investigation.

Keywords: Variational methods; Stochastic processes; Lagrangian densities,
Deformed derivatives.

\section{Some Mathematical Aspects and Comments on the use of Structural Derivatives}

\textbf{Conformable Derivative}

Recently, a promising new definition of local deformed derivative,
called conformable fractional derivative, has been proposed by the
authors in Ref. \citep{new definition} that preserve classical properties
and is given by 
\begin{equation}
T_{\alpha}f(t)=\lim_{\epsilon\rightarrow0}\frac{f(t+\epsilon t^{1-\alpha})-f(t)}{\epsilon}.
\end{equation}

If the function is differentiable in a classical sense, the definition
above yields 
\begin{equation}
T_{\alpha}f(t)=t^{1-\alpha}\frac{df(t)}{dt}.\label{eq:Differentiable}
\end{equation}

Changing the variable $t\rightarrow1+\frac{x}{l_{0}}$, we should
write (\ref{eq:Differentiable}) as $l_{0}\left(1+\frac{x}{l_{0}}\right)^{1-\alpha}\frac{d}{dx}f$,
that is nothing but the Hausdorff derivative up to a constant and
valid for differentiable functions.

For $t_{0}\neq0,$ an interval form of conformable derivative, the
left conformable derivative is 
\begin{equation}
T_{\alpha}^{a}f(t)=\lim_{\epsilon\rightarrow0}\frac{f(t+\epsilon(t-a)^{1-\alpha})-f(t)}{\epsilon},
\end{equation}

The use of deformed-operators was there also justified based on our
proposition that there exists an intimate relationship between dissipation,
coarse-grained media and the some limit scale of energy for the interactions
\citep{Jap-Weber-Sotolongo,EPL Weber}. Since we are dealing with
open systems, as commented in Ref. \citep{Nosso On conection2015}
, the particles are indeed dressed particles or quasi-particles that
exchange energy with other particles and the environment. Depending
on the energy scale an interaction may change the geometry of space\textendash time,
disturbing it at the level of its topology. A system composed by particles
and the surrounding environment may be considered nonconservative
due to the possible energy exchange. This energy exchange may be the
responsible for the resulting non-integer dimension of space\textendash time,
giving rise then to a coarse-grained medium. This is quite reasonable
since, even standard field theory, comes across a granularity in the
limit of Planck scale. So, some effective limit may also exist in
such a way that it should be necessary to consider a coarse-grained
space\textendash time for the description of the dynamics for the
system, in this scale. Also, another perspective that may be proposed
is the previous existence of a nonstandard geometry, e.g., near a
cosmological black hole or even in the space nearby a pair creation,
that imposes a coarse-grained view to the dynamics of the open system.
Here, we argue that deformed-derivatives allows us to describe and
emulate this kind of dynamics without explicit many-body, dissipation
or geometrical terms in the dynamical governing equations. In some
way, the formalism proposed here may yield an effective theory, with
some statistical average without imposing any specific nonstandard
statistics. So, deformed derivatives may be the tools that could describe,
in a softer way, connections between coarse-grained medium and dissipation
at a certain energy scale.

Also, we indicate that one relevant applicability of our formalism,
that concerns position-dependent systems \citep{Monteiro-Nobre-2013}
(see also Ref. \citep{Habib-Revisiting} and references therein),
seems to be more adequate to describe the dynamics of many real complex
systems, where there could exist long-rage interactions, long-time
memories, anisotropy, certain symmetry breakdown, non-linear media,
etc.  

An important point to emphasize is that the paradigm we adopt is different
from the standard approach in the generalized statistical mechanics
context, where the modification of entropy definition leads to the
modification of algebra and consequently the derivative concept. We
adopt that the mapping to a continuous fractal space leads naturally
to the necessity of modifications in the derivatives, that we will
call deformed or metric derivatives \citep{Balankin-Towards a physics on fractals}.
The modifications of derivatives brings to a change in the algebra
involved, which in turn may conduct to a generalized statistical mechanics
with some adequate definition of entropy.

\section{Variational Approach with embedded derivatives}

Our problem here is to search for minimizers of a variational problem
with what we now refer to as structural (or deformed/metric) derivatives
embedded into the Lagrangian function $L.$ After the mapping into
the fractal continuum, $L$ will be a C$^{2}$-function with respect
to all its arguments.

Remarks: (i)We consider a fractional variational problem which involves
local structural-derivatives called as Hausdorff that is in some sense
equivalent to the conformable derivative. Here we extend the previously
treated problem in Ref. \citep{Nosso variacional}, generalizing for
Lagrangian which will depend also on higher-order structural-derivatives.

(iii) We assume that $0<\alpha<1$.

(iv) Here, we adopt the Option 3 in Ref. \citep{Nosso variacional}:
Usual integral, $\delta$usual, structural-derivatives embedded, similar
to Ref. \citep{Atanackovic-Variational}, but here with local deformed
or metric derivatives.

We know that deformed-kernels used here can be replaced with other
kernels, resulting in a general variational calculus, as in Ref. \citep{Agrawal}.

Now, we are ready to set up the process:

In this case, we consider the fractional action $J[y]=\int_{a}^{b}L(x,y,D_{x}^{1}y,{}_{a}D_{x}^{\alpha}y,{}_{a}D_{x}^{\alpha}(D_{x}^{1}y))dx$
and the usual $\delta-$variational processes. Note the interval $[a,b]$
in action functional and the interval form of the conformable derivative
{[}Ref.{]}.

To derive the extended version of the Euler\textendash Lagrange equation
let us introduce the following $\alpha-$deformed functional action

We shall find the condition such that $J[y]$ has a local minimum. 

To do so, we consider the new fractional functional depending on the
parameter $\varepsilon$ .

Consider for the variable $y(x):$

$y(x)=y^{*}(x)+\varepsilon\eta(x);$ $y^{*}(x)$ is the objective
function, and $\eta(a)=\eta(b)=0$, $\varepsilon$ is a the parameter.

So, applying the deformed derivatives and the integer one, we obtain:

$_{a}D_{x}^{\alpha}y(x)={}_{a}D_{x}^{\alpha}y^{*}(x)+\varepsilon{}_{a}D_{x}^{\alpha}\eta(x).$

$D_{x}^{1}y(x)=D_{x}^{1}y^{*}(x)+\varepsilon D_{x}^{1}\eta(x).$

$_{a}D_{x}^{\alpha}(D_{x}^{1}y(x))={}_{a}D_{x}^{\alpha}(D_{x}^{1}y^{*}(x))+\varepsilon{}_{a}D_{x}^{\alpha}(D_{x}^{1}\eta(x)).$
with $_{a}D_{x}^{\alpha}\eta=(x-a)^{1-\alpha}\frac{d\eta}{dx},$ $D_{x}^{1}=\frac{dx}{dt}.$

Using the chain rule and the well known $\delta-$variational processes
relative to the $\varepsilon$ parameter, we can write

\[
\delta_{\varepsilon}L=\frac{\partial L}{\partial y}\eta+\frac{\partial L}{\partial(D_{x}^{1}y)}D_{x}^{1}\eta+\frac{\partial L}{\partial(_{a}D_{x}^{\alpha}y)}D_{x}^{\alpha}\eta+\frac{\partial L}{\partial(_{a}D_{x}^{\alpha}(D_{x}^{1}y))}{}_{a}D_{x}^{\alpha}((D_{x}^{1}\eta(x))),
\]

Since integration by part holds with deformed integral, similarly to the
integer case and, using that the usual transversality condition for an
extreme value, one obtains that$\delta_{\varepsilon}J=0$ implies that

The resulting $E-L$ equations are:

\begin{equation}
\frac{\partial L}{\partial y}-D_{x}^{1}\left(\frac{\partial L}{\partial(D_{x}^{1}y)}\right)-D_{x}^{1}[(x-a)^{1-\alpha}\frac{\partial L}{\partial(_{a}D_{x}^{\alpha}y)}]+D_{x}^{2}[(x-a)^{1-\alpha}\frac{\partial L}{\partial(_{a}D_{x}^{\alpha}(D_{x}^{1}y))}]=0,\label{eq:E-L Deformed High Order}
\end{equation}

\section{Relativistic, independent fields}

Now, we can proceed to pursue equivalent approaches to field theory,
based on independent relativistic fields.

Here, $\phi=\widetilde{\phi}+\epsilon_{1}^{\mu}\delta\phi,$ 

\[
\partial_{\mu}\phi=\partial_{\mu}\widetilde{\phi}+\epsilon_{1}^{\mu}\partial_{\mu}\delta\phi,
\]

Here,$\delta\phi,\delta\psi$ are arbitrary,$\widetilde{\phi},$$\widetilde{\psi}$are
the objective fields and $\mu=0,1,2,3,$ following the index spatial-temporal
derivative, $\partial_{\mu}.$

With deformed standard derivative, we usually consider the fractional
action

\[
S=\int dt\int d^{3}x\mathcal{L}(\phi,\partial_{\mu}\phi,\partial_{\mu}^{\alpha_{\lambda}}\phi,x^{\mu}),
\]
with the usual $\delta-$ process

\[
\delta_{\epsilon}\mathcal{L=\frac{\partial\mathcal{L}}{\partial\phi}\delta\phi}+\frac{\partial\mathcal{L}}{\partial\partial_{\mu}\phi}\delta\partial_{\mu}\phi+\frac{\partial\mathcal{L}}{\partial\partial_{\mu}^{\alpha_{\lambda}}\phi}\delta\partial_{\mu}^{\alpha_{\lambda}}\phi.
\]

$\partial_{\mu}\rightarrow\partial_{\mu}^{\alpha_{\lambda}}$, $\lambda=0,1,2,3;$
$\mathcal{L}(\phi,\partial_{\mu}\phi,\partial_{\mu}^{\alpha_{\lambda}}\phi,x^{\mu})$.

For the option 3 approach on our recent article Ref. \citep{Nosso variacional},
we have: 
\begin{eqnarray}
\frac{\partial\mathcal{L}}{\partial\phi}-\partial_{\mu}[\frac{\partial\mathcal{L}}{\partial(\partial_{\mu}\phi)}]-\partial_{\mu}[(x^{\mu})^{1-\alpha_{\lambda}}\frac{\partial\mathcal{L}}{\partial(\partial_{\mu}^{\alpha_{\lambda}}\phi)}] & = & 0.\label{eq:Field}
\end{eqnarray}

In the case of higher-order derivative, $\mathcal{L}(\phi,\partial_{\mu}\phi,\partial_{\mu}^{\alpha_{\lambda}}(\partial_{\mu}\phi),\partial_{\mu}^{\alpha_{\lambda}}\phi,x^{\mu})$,
it can be shown that the E-L equation result as

\[
\frac{\partial\mathcal{L}}{\partial\phi}-\partial_{\mu}[\frac{\partial\mathcal{L}}{\partial(\partial_{\mu}\phi)}]-\partial_{\mu}[(x^{\mu})^{1-\alpha_{\lambda}}\frac{\partial\mathcal{L}}{\partial(\partial_{\mu}^{\alpha_{\lambda}}\phi)}]+\partial_{\mu}^{2}[(x^{\mu})^{1-\alpha_{\lambda}}\frac{\partial\mathcal{L}}{\partial(\partial_{\mu}^{\alpha_{\lambda}}(\partial_{\mu}\phi))}]=0.
\]

Also, the derivatives can be considered in the intervalar form, $_{a}\partial_{\mu}\phi$,
for relativistic fields.

\section{Applications:}

In this Section, we apply our formalism to diverse physical problems.

\subsection{Dissipative Forces}

For details on obstruction to standard variational principles, the
reader is referred to Ref. \citep{Cresson}.

Here, starting off with a deformed Lagrangian and to gain some insight, we apply
the formalism to a simple case. We will show in the sequel that
dissipative systems can be treated with our formalism. 

Consider the Lagrangian of a particle with mass $m,$ submitted to
a position-dependent potential $U$:

$L=\frac{1}{2}m\overset{.}{(x)^{2}}-U(x)-\frac{1}{2}\gamma(_{a}D_{t}^{1/2}x)^{2},$
where the $\gamma$ is some parameter whose the physical meaning will
appear forward.

The corresponding deformed $E-L$ equation is 

\[
\frac{\partial L}{\partial x}-\frac{d}{dt}\frac{\partial L}{\partial D_{t}^{1}x}-\frac{d}{dt}[(t-a)^{1/2}\frac{\partial L}{\partial_{a}D_{t}^{1/2}x}]=0,
\]
 and leads to the movement equation

\begin{eqnarray}
m\frac{d^{2}x}{dt^{2}}+\frac{dU(x)}{dx}+\gamma\frac{d}{dt}[(t-a)\frac{dx}{dt}] & = & 0,
\end{eqnarray}
that can be rewritten as 
\[
m\frac{d^{2}x}{dt^{2}}+\frac{dU(x)}{dx}+\gamma[\frac{dx}{dt}+(t-a)\frac{d^{2}x}{dt^{2}}]=0.
\]

We now follow the limiting procedure indicated in Ref.
\citep{RW} with FC and in Ref. \citep{Matheus-Torres-Variat-Conformable},
with conformable derivatives. In the limit $a\rightarrow b\Longrightarrow(t-a)\rightarrow0,$
the resulting equation is 

\[
m\frac{d^{2}x}{dt^{2}}+\frac{dU(x)}{dx}+\gamma\frac{dx}{dt}=0.
\]
 Same result in Ref. \citep{Matheus-Torres-Variat-Conformable}, but
without any redefinition for the Lagrangian.

The equation stated above is a standard equation with the presence
of friction, showing the appearance of dissipation.

So, the results are evidently consistent with the classical Newtonian
mechanics with dissipation.

\subsection{Langevin Equation}

We shall show here that some the results of Ref. \citep{Metzler- Nature Rep}
for under-damped scaled Brownian motion can also be obtained by the
deformed variational procedure.

To this aim, let us now consider the Lagrangian:

\[
L=\frac{1}{2}m\overset{.}{(x)^{2}}-U(x)-\frac{1}{2}\gamma(_{a}D_{t}^{1/2}x)^{2}-\sqrt{2D(t)}\gamma(t)x(t)\zeta(t),
\]
where $\zeta(t)$ is a Gaussian noise \citep{Metzler- Nature Rep}.
Here we consider the time dependent diffusion coefficient $D(t)$
and the time dependent damping coefficient $\gamma(t).$

The E-L equation results as

\[
m\frac{d^{2}x}{dt^{2}}+\frac{dU(x)}{dx}+\left\{ \frac{d\gamma(t)}{dt}(t-a)\frac{dx}{dt}+\gamma(t)[\frac{dx}{dt}+(t-a)\frac{d^{2}x}{dt^{2}}]\right\} =\sqrt{2D(t)}\gamma(t)\zeta(t).
\]

In the limit $a\rightarrow b,$ we obtain the Langevin equation

\begin{equation}
m\frac{d^{2}x}{dt^{2}}+\frac{dU(x)}{dx}+\gamma(t)\frac{dx}{dt}=\sqrt{2D(t)}\gamma(t)\zeta(t).\label{eq:Langevin}
\end{equation}

Now, following again Ref.\citep{Metzler- Nature Rep}, if we consider
the case of relevant parameters and potential as 
\begin{eqnarray*}
\gamma(t) & = & \gamma_{0}(1+\frac{t}{\tau})^{\alpha-1}\\
D(t) & = & D_{0}(1+\frac{t}{\tau})^{\alpha-1}\\
U(x) & = & 0,
\end{eqnarray*}

The Langevin equation now reads 

\[
m\frac{d^{2}x}{dt^{2}}+\gamma_{0}(1+\frac{t}{\tau})^{\alpha-1}\frac{dx}{dt}=\sqrt{2D_{0}}\gamma_{0}(1+\frac{t}{\tau})^{\frac{3}{2}(\alpha-1)}\zeta(t).
\]

That is the result in the Ref. \citep{Metzler- Nature Rep} for underdamped
scaled Brownian motion.

\subsection{Abraham-Lorentz Lagrangian}

We now proceed with application related to a Abraham-Lorentz force.

To pursue this objective, that is, to obtain the back-reaction equation,
let us now consider the Lagrangian 

\[
L=\frac{1}{2}m\overset{.}{(x)^{2}}-U(x)+\frac{e^{2}}{6c^{3}}([_{a}D_{t}^{1/2}(\frac{dx}{dt})]^{2}.
\]

Using now eq.(\ref{eq:E-L Deformed High Order}), we obtain the E-L
equation as 

\[
m\frac{d^{2}x}{dt^{2}}-\frac{dU(x)}{dt}+\frac{2e^{2}}{6c^{3}}\frac{d}{dt}\{\frac{d}{dt}[(t-a)\frac{d^{2}x}{dt^{2}}]\}=0,
\]
 that can be rewritten as

\[
m\frac{d^{2}x}{dt^{2}}-\frac{dU(x)}{dt}+\frac{2e^{2}}{6c^{3}}\{\frac{d^{3}x}{dt^{3}}+\frac{d^{3}x}{dt^{3}}+(t-a)\frac{d^{4}x}{dt^{4}}\}=0.
\]

Taking the limit $a\rightarrow b\Longrightarrow(t-a)\rightarrow0,$
the resultant equation is the one with the Abraham Lorentz term for
the radiation reaction,

\[
m\frac{d^{2}x}{dt^{2}}-\frac{dU(x)}{dt}+\frac{2e^{2}}{3c^{3}}\frac{d^{3}x}{dt^{3}}=0.
\]

\subsection{Adapted Galley Method }

Here, we make an attempt to apply a simplified form of the method
due to Galley, in Ref. \citep{Galley}.

Following Ref. \citep{Nosso variacional}, considering the Lagrangian
and the action functional as 
\begin{eqnarray*}
L & = & L(x,y,D_{x}^{1}y,_{a}D_{x}^{\alpha}y,_{a}D_{x}^{\alpha}(D_{x}^{1}y),D_{x}^{1}z,_{a}D_{x}^{\alpha}z),\\
J[L] & = & \int_{a}^{b}L(x,y,D_{x}^{1}y,_{a}D_{x}^{\alpha}y,_{a}D_{x}^{\alpha}(D_{x}^{1}y),D_{x}^{1}z,_{a}D_{x}^{\alpha}z)dx.
\end{eqnarray*}

Here the variable $z$ can be considered as the duplicated variable
in the context of Galley formalism. So, there are two E-L equations

\begin{equation}
\frac{\partial L}{\partial y}-D_{x}^{1}\left(\frac{\partial L}{\partial(D_{x}^{1}y)}\right)-D_{x}^{1}[(x-a)^{1-\alpha}\frac{\partial L}{\partial_{a}D_{x}^{\alpha}y}]+D_{x}^{2}[(x-a)^{1-\alpha}\frac{\partial L}{\partial(_{a}D_{x}^{\alpha}(D_{x}^{1}y))}]=0,
\end{equation}

\begin{equation}
\frac{\partial L}{\partial z}-D_{x}^{1}\left(\frac{\partial L}{\partial(D_{x}^{1}z)}\right)-D_{x}^{1}((x-a)^{1-\gamma}\frac{\partial L}{\partial(_{a}D_{x}^{\alpha}z)})=0.
\end{equation}

Considering now the Lagrangian as 
\[
L=L_{x}-L_{z}+L_{xz}=\frac{1}{2}m\overset{.}{(x)^{2}}-U(x)-\frac{1}{2}m\overset{.}{(z)}+U(z)+\frac{2e^{2}}{3c^{3}}[_{a}D_{t}^{\alpha}(\frac{dx}{dt})][_{a}D_{t}^{\alpha}z],
\]
we can can write, using the with standard notation for first order
derivative $D_{x}^{1}=\frac{dx}{dt}$, the E-L Equations as

\begin{eqnarray*}
m\frac{d^{2}x}{dt^{2}}-\frac{dU(x)}{dx}+\frac{2e^{2}}{6c^{3}}\frac{d}{dt}\{\frac{d}{dt}[(t-a)^{1-\alpha}{}_{a}D_{t}^{\alpha}z]\} & = & 0,\\
-m\frac{d^{2}z}{dt^{2}}+\frac{dU(z)}{dx}-\frac{2e^{2}}{6c^{3}}\frac{d}{dt}[(t-a)^{1-\alpha}{}_{a}D_{t}^{\alpha}(\frac{dx}{dt})] & = & 0.
\end{eqnarray*}

or 
\begin{eqnarray*}
m\frac{d^{2}x}{dt^{2}}-\frac{dU(x)}{dx}+\frac{2e^{2}}{3c^{3}}\frac{d}{dt}\{(1-2\alpha)(2-2\alpha)(t-a)^{-2\alpha}\frac{dz}{dt}+(2-2\alpha)(t-a)^{1-2\alpha}\frac{d^{2}z}{dt^{2}}+(1-\alpha)(t-a)^{-\alpha}\frac{d^{2}z}{dt^{2}}+(t-a)^{1-\alpha}\frac{d^{3}z}{dt^{3}}\} & = & 0,\\
-m\frac{d^{2}z}{dt^{2}}+\frac{dU(z)}{dx}-\frac{2e^{2}}{3c^{3}}[(2-2\alpha)(t-a)^{1-2\alpha}\frac{d^{2}x}{dt^{2}}+(t-a)^{2-2\alpha}\frac{d^{3}x}{dt^{3}}] & = & 0.
\end{eqnarray*}

Considering now the limit $\alpha\rightarrow1$ and for physical meaning,
collapse the variables $x$ and $z$ into one, that is $x=z$.

So, we obtain a unique equation 

\[
m\frac{d^{2}x}{dt^{2}}-\frac{dU(x)}{dx}+\frac{2e^{2}}{3c^{3}}\frac{d^{3}x}{dt^{3}}=0.
\]

This may indicate the equivalence between Galley's method for integer
derivatives and the simple one with structural derivatives, confirming
that the variational calculus with structural derivatives (here, the
conformable derivative is in its ``intervalar'' form) treats the
hidden degrees of freedom in a simple way.

\subsection{Reaction-Convection-Diffusion Equation}

It is known \citep{Cresson} that many physical systems may be modeled
by the convection-diffusion equation. When the laws of thermodynamics
predict a different behavior for macroscopic systems, compared to
the behavior of individual molecules, the concept of energy dissipation
may appears. This is the case of systems such as fluid particles,
e.g., in the fields of confined and free-surface flows \citep{Violeau}.
We here claim that the correct treatment of dissipative forces is
the deformed Lagrangian methods such as that with the use of intervalar
form of conformable derivatives.

Consider the Lagrangian density Ref. \citep{Cresson}. As in the indicated
reference, $U(t,x)$ is some field concentration or field density
that is space and time dependent. The tensor $K$ represents the diffusivity,
$f$ is the source term and $\gamma\in\mathbb{R^{\mbox{d}}}$ is some
flow velocity.

$L=L(t,x,U(t,x),\nabla U(t,x),_{a}D_{t}^{\alpha}U,_{x_{a}}\nabla_{x}^{\alpha}U)=f(t,x)U(t,x)-\frac{1}{2}\beta U(t,x)^{2}+\frac{1}{2}(_{a}D_{t}^{1/2}U)^{2}+\frac{1}{2}(\gamma\times{}_{x_{a}}\nabla_{x}^{1/2}U).(_{x_{a}}\nabla_{x}^{1/2}U)-\frac{1}{2}(K.\nabla U(t,x)).(\nabla U(t,x)),$
where $x=\left(\begin{array}{c}
x_{1}\\
x_{2}\\
x_{3}
\end{array}\right),$

$\nabla U(t,x)=\left(\begin{array}{c}
\frac{\partial U}{\partial x_{1}}\\
\frac{\partial U}{\partial x_{2}}\\
\frac{\partial U}{\partial x_{3}}
\end{array}\right),$ $_{x_{a}}\nabla_{x}^{\alpha}U=\left(\begin{array}{c}
_{x_{a_{1}}}D_{x_{1}}^{\alpha}U\\
_{x_{a_{2}}}D_{x_{2}}^{\alpha}U\\
_{x_{a_{3}}}D_{x_{3}}^{\alpha}U
\end{array}\right)=\left(\begin{array}{c}
(x_{1}-x_{a_{1}})^{1-\alpha}\frac{\partial U}{\partial x_{1}}\\
(x_{2}-x_{a_{2}})^{1-\alpha}\frac{\partial U}{\partial x_{2}}\\
(x_{3}-x_{a_{3}})^{1-\alpha}\frac{\partial U}{\partial x_{3}}
\end{array}\right).$

\begin{eqnarray*}
\frac{\partial L}{\partial U} & = & f(t,x)-\beta U\\
(t-a)^{1/2}\frac{\partial L}{\partial(_{a}D_{t}^{1/2}U)} & = & (t-a)^{1/2}{}_{a}D_{t}^{1/2}U=(t-a)\stackrel{\cdotp}{U}\Longrightarrow-\frac{\partial}{\partial t}[(t-a)\stackrel{\cdotp}{U}]=-[\stackrel{\cdotp}{U}+(t-a)\stackrel{\cdot\cdot}{U}].
\end{eqnarray*}

In the limit $a\rightarrow b,$ the term is $-\stackrel{\cdotp}{U}$.

Also,

\[
-\nabla[(x-x_{a})^{1/2}.\frac{\partial L}{\partial{}_{x_{a}}\nabla_{x}^{1/2}U}]=-\nabla[(x-x_{a})^{1/2}.\gamma.({}_{x_{a}}\nabla_{x}^{1/2}U)]=-[\gamma.(\nabla U(t,x))+\gamma(x-x_{a})\triangle^{2}U].
\]

Taking the limit $a\rightarrow b,$we obtain the term 
\[
-\gamma.(\nabla U(t,x)).
\]

The last parcel in E-L equation is

\[
-\nabla.[\frac{\partial L}{\partial(\nabla U(t,x))}]=\nabla.(K.\nabla U).
\]

The EL equation gives the reaction-convection-diffusion Equation \citep{Cresson},
that have not been obtained from a variational principle with standard
derivatives.

\[
\frac{\partial U}{\partial t}+\gamma.(\nabla U(t,x))-\nabla.(K.\nabla U)+\beta U=f(x,t).
\]

\subsection{Linear Fokker-Planck Equation}

Consider now the Lagrangian density

$L=\frac{1}{2}({}_{a}D_{t}^{1/2}P)^{2}-\frac{1}{2}(D.\nabla P).\nabla P-\frac{1}{2}(\nabla f(x)).P^{2}+\frac{1}{2}[f(x)\times({}_{x_{a}}\nabla_{x}^{1/2}P)]\times({}_{x_{a}}\nabla_{x}^{1/2}P).$

The resulting EL equation gives

\[
\frac{\partial P}{\partial t}=-\nabla.[f(x)\times P(x,t)]+D.\triangle P.
\]

In one dimension, it gives the linear Fokker-Planck
equation

\[
\frac{\partial P}{\partial t}=-\frac{\partial}{\partial x}.[f(x).P(x,t)]+D\frac{\partial^{2}P}{\partial x^{2}}.
\]

\subsection{Non-linear Fokker-Planck Equation}

Consider now the Lagrangian $L$, with $P$ as an event probability
in the statistical physics context of Fokker-Plank equations,

\[
L=\frac{1}{2}({}_{a}D_{t}^{1/2}P)^{2}-\frac{1}{2}[(D.\nabla P).\nabla P]P^{\mu-1}-\frac{1}{2}(\nabla f(x)).P^{2}+\frac{1}{2}[f(x)\times({}_{x_{a}}\nabla_{x}^{1/2}P)]\times({}_{x_{a}}\nabla_{x}^{1/2}P).
\]

The resultant equation, in one dimensional form, is 

\[
\frac{\partial P}{\partial t}=-\frac{\partial}{\partial x}.[f(x).P(x,t)]+D\frac{\partial}{\partial x}[P^{\mu-1}\frac{\partial P}{\partial x}]+\frac{1}{2}(\mu-1)D(\frac{\partial P}{\partial x})^{2}P^{\mu-2}.
\]

This is similar to the proposed equation in Ref. \citep{Nobre Curado-Fokker-Planck}
but here with the additional last term.

In the case of $\mu-1=\varepsilon\approx1,$the lest term drops and
the resultant equation is identical of Ref. \citep{Nobre Curado-Fokker-Planck},
but with the condition of smooth non-linearity indicated.

Another type of non linear Fokker-Planck equation, can result from
the Lagrangian

\[
L=\frac{1}{2}({}_{a}D_{t}^{1/2}P)(_{a}D_{t}^{1/2}P^{\mu})-\frac{1}{2}[(D.\nabla P).\nabla P]P^{\nu-1}-\frac{\nabla f(x).P^{\mu+1}}{\mu+1}+\frac{1}{2}[f(x)\times({}_{x_{a}}\nabla_{x}^{1/2}P)]\times({}_{x_{a}}\nabla_{x}^{1/2}P)P^{\mu}.
\]

The resulting dynamical equation, in one dimension, from EL equation
is

\[
\frac{\partial P^{\mu}}{\partial t}=-\frac{\partial}{\partial x}.[f(x).(P(x,t))^{\mu}]+D\frac{\partial}{\partial x}[P^{\nu-1}\frac{\partial P}{\partial x}]+\frac{1}{2}(\nu-1)D(\frac{\partial P}{\partial x})^{2}P^{\nu-2}.
\]

The equation above is similar to Ref.\citep{Tsallis 1996}. If we
put $\nu=1,$the equation above coincide with that of Ref. \citep{Tsallis 1996},
for this value of $\nu.$

\subsection{KDV Equation - Lagrangian formulation without auxiliary potential
$\psi$}

Consider now a Lagrangian described by three terms as follows:

$L=L_{1}+L_{2}+L_{3},$ with

\begin{eqnarray}
L_{1} & = & \frac{1}{4}((_{x_{a}}D_{x}^{1/2}(\frac{\partial\phi}{\partial x}))^{2},\\
L_{2} & = & -\frac{1}{2}(_{a}D_{t}^{1/2}\phi),\\
L_{3} & = & 3\phi(_{x_{a}}D_{x}^{1/2}\phi)^{2}.
\end{eqnarray}
The resulting E-L is nothing but the the KDV equation:
\begin{equation}
\frac{\partial\phi}{\partial t}+\frac{\partial^{3}\phi}{\partial x^{3}}-6\phi\frac{\partial\phi}{\partial x}=0.
\end{equation}

Note that no auxiliary potential was introduced.

Now, by considering in the sequence some deformations of the previous
Lagrangian, we will show that some Deformed KDV equation emerges.
Consider the Lagrangian terms

\begin{eqnarray}
L_{1} & = & \frac{1}{4}((_{x_{a}}D_{x}^{1/2}(\frac{\partial\phi}{\partial x}))^{2},\\
L_{2} & = & -\frac{1}{2}(_{a}D_{t}^{1/2}\phi)(_{a}D_{t}^{1/2}\phi^{\mu}),\\
L_{3} & = & 3\phi^{\nu}(_{x_{a}}D_{x}^{1/2}\phi)^{2}.
\end{eqnarray}

The resulting deformed E-L is the deformed-KDV equation:
\begin{equation}
\frac{\partial\phi^{\mu}}{\partial t}+\frac{\partial^{3}\phi}{\partial x^{3}}-6\phi^{\nu}\frac{\partial\phi}{\partial x}=0.
\end{equation}

\subsection{LLG Equation - Lagrangian Formulation with Conformable Derivatives}

Following steps in Ref. \citep{Bose-Trimper}, we will show that the
Landau-Lifshits-Gilbert equation could also be obtained by our approach.

To this intend, consider the Lagrangian:

$L=2A_{\nu}\overset{\cdot}{m}_{\nu}-\frac{1}{2}\kappa c(_{a}D_{t}^{1/2}m_{\beta})^{2}-H_{eff_{\beta}}m_{\beta};$

$H_{eff_{\beta}}=H_{\beta}-\sigma\overset{\cdot}{m}_{\beta}.$

E-L LLG(component $\beta$):

$[\overrightarrow{m}\wedge(\nabla_{m}\wedge\overrightarrow{A}(\overrightarrow{m}))]_{\beta}-\overset{\rightarrow}{H}_{eff_{\beta}}+\kappa c(\overset{\cdot}{\overrightarrow{m}})_{\beta}=0.$

With \citep{Bose-Trimper} $(\nabla_{m}\wedge\overrightarrow{A}(\overrightarrow{m}))=g\overrightarrow{m},$
and $m^{2}=1,$ the equation above can be rewritten as the LLG equation.
That is,

\begin{equation}
\frac{\partial\overrightarrow{m}}{\partial t}=\frac{1}{g}[\overrightarrow{m}\wedge\overset{\rightarrow}{H_{eff}}]-\frac{\kappa c}{g}[\overrightarrow{m}\wedge\overset{\cdot}{\overrightarrow{m}}].
\end{equation}

With $\gamma=-\frac{1}{g},$ $\alpha=-\frac{\kappa c}{g}=\kappa c\gamma,$
the equation can be converted to usual LLG form

\begin{eqnarray}
\frac{\partial\overrightarrow{m}}{\partial t} & = & -\frac{\gamma}{(1+\alpha^{2})}[\overrightarrow{m}\wedge\overset{\rightarrow}{H_{eff}}]-\frac{\alpha\gamma}{(1+\alpha^{2})}[\overrightarrow{m}\wedge(\overrightarrow{m}\wedge\overset{\rightarrow}{H_{eff}})].
\end{eqnarray}

\subsection{Caldirola-Kanai Hamiltonian by $\lambda-exp$ Metric Derivative\label{sec:Caldirola-Kanai-Hamiltonian-by}}

\textbf{New $\lambda-$Exponential Metric (or Conformable) Derivative}

Lets us define a new metric derivative, that will be related, as will
be shown, to Caldirola-Kanai Hamiltonian

\[
\mathcal{D}_{t}^{\lambda}(f)(t)=\lim_{\epsilon\rightarrow0}\frac{f(t+\epsilon e^{-\lambda t})-f(t)}{\epsilon},
\]

With a change in variable $\epsilon e^{-\lambda t}=\epsilon',$we
can write the $\lambda-exp$ metric derivative as

\begin{equation}
\mathcal{D}_{t}^{\lambda}(f)(t)=e^{-\lambda t}\frac{df(t)}{dt}.\label{eq:Exponent Metric deriv}
\end{equation}
 We can see that the above form of deformed derivative, for differentiable
functions, has the same mathematical structure of the generic deformed
metric derivative, so we can use the variational approach indicated
in our recent work \citep{Nosso variacional} , to obtain interesting
result related to dissipative systems, particularly interesting is
the Caldirola-Kanai Hamiltonian.

Consider the $\lambda-exp$ metric derivative given by eq.(\ref{eq:Exponent Metric deriv}).
We can use the variational approach proposed and write for the E-L
equation as

\[
\frac{\partial L}{\partial q}-\frac{d}{dt}(e^{-\frac{\lambda}{2}t}\frac{\partial L}{\partial D_{t}^{\lambda}q})=0,
\]

With a Hamiltonian \citep{Nosso variacional} $H=H(p^{\lambda},q,t)\equiv p^{\lambda}(D_{t}^{\lambda}q)-L,$
where $p^{\lambda}=\frac{\partial L}{\partial(D_{t}^{\lambda}q)}.$
Here $q$ is some generalized coordinate and have not to be confused
with the entropic parameter \citep{Nosso variacional}. 

Now, consider a ``quasi-particle'', with a Lagrangian given by $L=\frac{1}{2}m(D_{t}^{\lambda}q)^{2}+V(t),$where
$V(t)$ is a time dependent ``potential-like'' term given in by
similarity with an harmonic oscillator with an elastic time-dependent
constant. So, the time-dependent potential can be written as $V(t)=\frac{1}{2}k(t)q^{2}=\frac{1}{2}me^{\lambda t}\omega_{0}q^{2}.$

Withe the above expressions, we can write for the Hamiltonian $H=p^{\lambda}(D_{t}^{\lambda}q)-L=\frac{1}{2}m(D_{t}^{\lambda}q)^{2}+\frac{1}{2}me^{\lambda t}\omega_{0}q^{2}.$

Now, the $\lambda-exp$ metric derivative of the generalized coordinate
is 
\[
D_{t}^{\lambda}q=e^{-\frac{\lambda}{2}t}\frac{dq}{dt}=e^{-\frac{\lambda}{2}t}\frac{p}{m},
\]
where $p$ is the generalized moment. With this, the Hamiltonian can
be written as 

\begin{equation}
H=e^{-\lambda t}\frac{p^{2}}{2m}+\frac{1}{2}me^{\lambda t}\omega_{0}q^{2},\label{eq:CK Hamiltonian}
\end{equation}
 that is nothing but the Caldirola-Kanai Hamiltonian, with important
implications in the study of dissipative systems.

This is a strong indication that the metric involved is the responsible
for an apparent mass-time dependent term and that the approach in
\citep{Nosso variacional} is an adequate variational approach to
deal with dissipative systems.

\section{Conclusions and Outlook}

In conclusion, we have extended the approach in Ref. \citep{Nosso variacional}
and employed the variational calculus to obtain E-L equations with
deformed derivatives. We believe that with this formalism can set
up a systematic way to obtain nonstandard equations in several areas
of science, without an excessive heuristics. This can avoid to introduce
ad hoc fields and unnecessary suppositions about strange dynamics.

We have shown that our approach is suitable to describe, within the Lagrangian
formalism, stochastic systems, Langevin, Reaction- Convection-Diffusion
equation, Fokker-Planck equations, Abraham-Lorentz radiation reaction
force, KDV equation and so on. By building up Lagrangians with deformed
derivatives for several dynamical systems, we have shown that the
mathematical tool can be applied to the study and development of several
areas in science.

For the future works, we shall apply the approach to quantum systems,
by following, e.g., the canonical method, where the time derivative
of physical quantities can be obtained via a modified Poisson bracket
formalism, and then perform the quantization. We will also apply a
Hamilton formalism for dissipative systems using conformable derivatives.

The inclusion of higher-order derivatives could lead to an interesting
study of Podolsky-like systems and shall be the matter of a future publication.

\textbf{\bigskip{}
}

\end{document}